\documentclass[11pt,a4paper,oneside]{article}
\usepackage[T1]{fontenc}
\usepackage[utf8]{inputenc}
\usepackage{geometry}  
\usepackage[american]{babel}
\usepackage[babel]{csquotes}
\usepackage{amssymb}
\usepackage{xpunctuate} 
\newcommand{\eg}[0]{e.\,g\xperiod}
\newcommand{\ie}[0]{i.\,e\xperiod}

\usepackage[
            style=numeric-comp,
            sorting=nyt,
            abbreviate=true,
            useprefix=true,
            giveninits=true,
            maxcitenames=3,
            maxbibnames=20,
            uniquename=init,
            sortcites=true,
            doi=true,
            isbn=false,
            url=false,
            eprint=false,
            backend=biber
                           ]{biblatex} 
  \renewbibmacro{in:}{}
  \DefineBibliographyStrings{english}{bibliography = {References}}
  \preto\fullcite{\AtNextCite{\defcounter{maxnames}{99}}}
  \AtEveryBibitem{\clearfield{month}} 
  \AtEveryCitekey{\clearfield{month}}

\usepackage{enumitem}
\usepackage[font=scriptsize]{caption}
\usepackage{graphicx}
\usepackage{booktabs} 
\usepackage{tabularx}  
  \newcolumntype{R}{>{\raggedleft\arraybackslash}X}
  \newcolumntype{L}{>{\raggedright\arraybackslash}X}
  \newcolumntype{C}{>{\centering\arraybackslash}X}

\usepackage{hyperref}
\hypersetup{
final,                     
pdftoolbar=true,          
pdfmenubar=true,          
pdftitle={Estimating saturation-dependent diffusion from 3D~micro-CT images using CNNs},    
pdfauthor={G\"arttner et al.},  
pdfnewwindow=true,         
colorlinks=true,           
linkcolor=[rgb]{0,0.3,0.3},           
citecolor=[rgb]{0,0.3,0.3},           
filecolor=[rgb]{0,0.3,0.3},            
urlcolor=[rgb]{0,0.3,0.3}              
}

\usepackage{mathtools}
\usepackage[OMLmathsfit]{isomath}

\usepackage{siunitx}                 
\usepackage{setspace}

\usepackage{dsfont}

\usepackage[symbol]{footmisc}

\newcommand{\IR}{\ensuremath\mathds{R}}
\renewcommand*{\vec}[1]{{\boldsymbol{#1}}}
\newcommand*{\grad}{\vec{\nabla}}                               
\renewcommand*{\div}{\vec{\nabla}\cdot}                          
\newcommand*{\dd}{\mathrm{d}} 
\usepackage{amsthm}
\usepackage{cleveref} 
  \crefformat{equation}{(#2#1#3)}
  \Crefformat{equation}{Equation~#2#1#3}
  \crefformat{figure}{Figure~#2#1#3}         \crefname{figure}{Figure}{Figures}
  \crefformat{section}{Section~#2#1#3}      \crefname{section}{Section}{Sections}
  \Crefformat{section}{Section~#2#1#3}      \Crefname{section}{Section}{Sections}
  \crefformat{table}{Table~#2#1#3}          \crefname{table}{Table}{Tables}
  \Crefformat{table}{Table~#2#1#3}          \Crefname{table}{Table}{Tables}

\newcommand*{\orcid}[1]{\href{https://orcid.org/#1}{\includegraphics[width=1em]{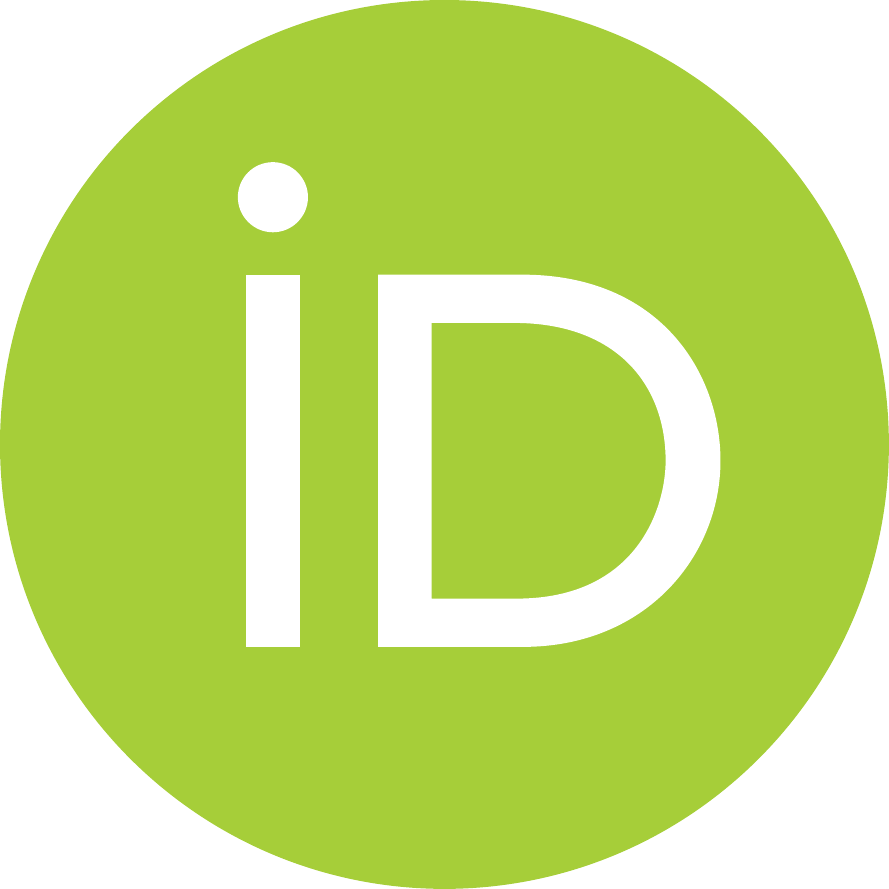}}}
\title{Estimating relative diffusion from 3D~micro-CT images using CNNs}
\author{Stephan G\"arttner\textsuperscript{1,*}\orcid{0000-0002-3488-0929},
Florian Frank\textsuperscript{1,2,*}\orcid{0000-0002-9756-1351},
Fabian Woller\textsuperscript{1}\orcid{0000-0001-5492-6819},
Andreas Meier\textsuperscript{1}\orcid{0000-0002-3135-9310}, 
\\
Nadja Ray\textsuperscript{1}\orcid{0000-0002-9596-953X}
\\
\normalsize\begin{tabular}{rl}
{}\\
\textsuperscript{1} & Friedrich-Alexander-Universität Erlangen-Nürnberg, \\
{} & Department~Mathematik, Cauerstra{\ss}e~11, 91058~Erlangen, Germany
\\
\textsuperscript{2} & Math2Market GmbH, Richard-Wagner-Stra{\ss}e~1, 67655~Kaiserslautern, Germany
\\
\end{tabular}
}
\date{\today}

\begin{document}
\setstretch{1.3}

\maketitle
\thispagestyle{empty}

\footnotetext{\textsuperscript{*} Corresponding authors. \emph{E-mail address:} \url{gaerttner@math.fau.de}, \url{frank@math.fau.de}.}



\begin{abstract}
In the past several years, convolutional neural networks (CNNs) have proven their capability to predict characteristic quantities in porous media research directly from pore-space geometries. Due to the frequently observed significant reduction in computation time in comparison to classical computational methods, bulk parameter prediction via CNNs is especially compelling, \eg for effective diffusion. While the current literature is mainly focused on fully saturated porous media, the partially saturated case is also of high interest. Due to the qualitatively different and more complex geometries of the domain available for diffusive transport present in this case, standard CNNs tend to lose robustness and accuracy with lower saturation rates. In this paper, we demonstrate the ability of CNNs to perform predictions of relative diffusion directly from full pore-space geometries. As such, our CNN conveniently fuses diffusion prediction and a~well-established morphological model which describes phase distributions in partially saturated porous media.
\end{abstract}

\section{Introduction}

Data-driven methods have proven to be powerful instruments for the solution of central tasks in digital rock physics, ranging from segmentation of pore-scale images~\cite{WANGReview} to the prediction of effective quantities such as porosity, permeability, and diffusion~\cite{graczyk2020predicting,CNNDiffusivityWu19} of pore-space geometries. Especially for the latter case, convolutional neural networks (CNNs) play a~dominant role. In~\cite{CNNDiffusivityWu19}, a~2D CNN was trained on artificially generated data-sets for diffusion prediction achieving superior accuracy in comparison to heuristic approaches such as the Bruggeman equation. Similar results were obtained in~\cite{WANG2020100035,CNNPrifling} for the 3D case on artificially generated geometries, underlining the enormous speed-up of several orders of magnitude in comparison to standard lattice Boltzmann diffusion solvers. 
\par
Existing literature concentrates on the consideration of fully saturated samples, in which the complete pore space is equally available for diffusive processes. In this paper, we investigate the capabilities of CNNs to perform accurate diffusion predictions on partially saturated samples in 3D. As such, diffusion is restricted to a certain subdomain of the full pore space. In numerical studies~\cite{DiffSatNum} as well as in experimental studies~\cite{DiffSatExp}, a~sample's relative diffusion is reported to vary significantly with saturation. However, in general, structure and connectivity of the pore space also influence relative diffusion. These parameters typically play an~important role in imbibition or drainage processes, the simulation of which requires efficient computations of the evolving relative diffusion.
\par
In our study, we assume the pore space to be filled with two different static fluid phases, where one is completely wetting (\eg water) and the other is fully non-wetting (gas), \ie developing zero-degree contact angles. Moreover, we assume only the wetting phase to allow for diffusive transport. As such, the medium's relative diffusion coincides with its absolute diffusion with respect to the wetting-phase domain. Consequently, this domain's geometry is of major importance to our study. It is expected to admit lower degrees of pore-connectivity and higher complexity with lower saturation, exacerbating the applicability of heuristic laws. To address realistic scenarios, we predict the relative diffusion of real rock micro-CT scans instead of artificially generated samples.
\par
In the literature, multiple methods are available to predict the distribution of two fluid phases in the pore space of a~porous medium, a~comparison of which is found in~\cite{POTMorphExp}. According to this study, lattice Boltzmann methods provide a~very good approximation to the experimentally measured phase distribution at the expense of large computational overheads, cf.~\cite{LBMPhaseDistGenty}. A different approach is posed by pore-morphology models (maximum spheres method) as elaborated in~\cite{MU201616,Silin11,HILPERT2001243,POTMorphExp}. In this purely geometric approach, balls are placed within the pore space, successively displacing the wetting fluid phase until a~desired saturation level is reached. As indicated in~\cite{MU201616}, the resulting algorithm can be implemented efficiently while still reflecting the main features of the related experiments, cf.~\cite{POTMorphExp}. Consequently, we make use of the second approach to distribute the non-wetting phase in the pore space of our CT samples.   
\par
In this study, we first demonstrate that standard CNNs trained on fully saturated pore spaces poorly generalize to partially saturated samples. More precisely, the prediction quality deteriorates monotonously with the saturation level. This behavior is expected due to the increasingly significant structural differences of the geometry available for diffusive transport. Our main aim then, is to demonstrate the capability of CNNs to predict relative diffusion without explicit precomputation of the phase distribution. 
\par
We note that for the similar problem of permeability prediction, deep neural networks were already successfully used to predict relative permeability from numerous precomputed and carefully-chosen sample properties, cf.~\cite{DNNRelPerm}. Contrary to this approach, we novelly train a~CNN to perform relative diffusion predictions directly from the full pore-space geometry. To do so, the network is trained on relative diffusion data computed from the explicit phase distribution as dictated by the morphological model. Our CNN therefore learns to implicitly reflect the influence of saturation on the target quantity. As such, for the first time, it combines the application of a~morphological model with the subsequent diffusion computation on the resulting geometry into a~monolithic data-driven approach. Therefore, our method is well-suited to simplify workflows in large-scale transport simulations involving local dynamic changes in saturation and/or porous matrix structure. In these scenarios, bulk parameters are to be steadily recomputed from reference volumes. This applies to quasi-static drainage processes as well as dissolution/precipitation scenarios, which both require repeated evaluations of current effective parameters like relative diffusion over time. 
\par
This paper is outlined as follows: In \cref{SEC:Datapreparation}, we describe the methodology used for generating a~suitable set of training data. This includes the morphological model for phase distribution in a~partially saturated pore space as well as the forward simulation to label the resulting samples with the corresponding diffusion value. \cref{SEC:MachineLearningModel} presents the architecture chosen for our CNN. In \cref{SEC:Evaluation}, we investigate the performance of our CNN to predict relative diffusion from full pore-space geometries. Therefore, we train on data which take the explicit phase distribution given by the morphological model into account. Finally, a~conclusion of our results is presented in~\cref{SEC:SummaryandConclusions}.

\section{Methodology and data preparation}
\label{SEC:Datapreparation}
In this section, we present the overall workflow, by which our training data set is generated. First, we provide an~overview of the sampling procedure, extracting suitable $100^3$-voxel training samples from a~larger micro-CT scan. Subsequently, the morphological model creating a~realistic phase distribution for a~given partial saturation value of the pore space is described. The resulting subdomain of the wetting fluid is then used to perform computations determining the sample's relative diffusion. Details on the forward simulation of our workflow as well as the CNN design are provided below. 

\begin{figure}[h!]
\centering
    \includegraphics[width=0.9\textwidth]{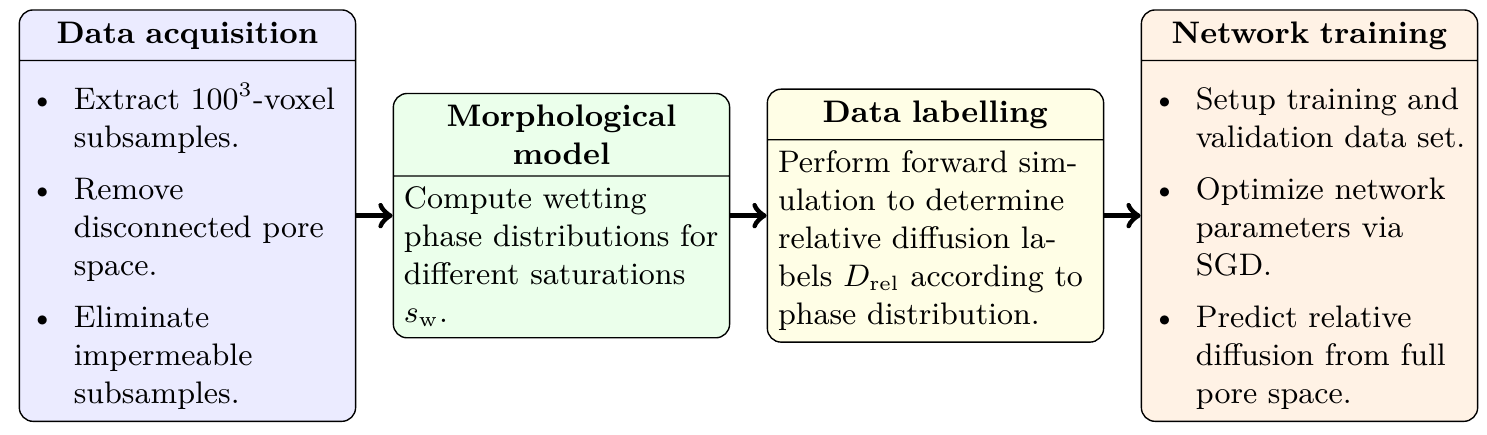}
    \caption{Overall workflow in flow chart representation. }
    \label{fig:FlowChart}
\end{figure}

\subsection{Sampling and preprocessing} 
\label{SEC:DataSampling}

In order to properly assess the prediction quality of our CNN on realistic geometries of partially saturated porous media, we prepare our training data set from segmented micro-CT scans of a~sandstone specimen, cf.~\cite{BentheimerDataSet}. As a~basis for this study, we make use of the data set generated in~\cite{GaerttnerCNNPerm} for the purpose of permeability predictions using CNNs. The related software as well as the data set are available in the package RTSPHEM~\cite{RTSPHEM}.
In the following, we outline the major processing steps.
\par
All training samples are derived from a~$1000^3$ binary voxel image of a~Bentheimer sandstone as provided in~\cite{BentheimerDataSet,BentheimerPublication}, featuring a~resolution of $\SI{2.25}{\micro m}$ per voxel. Due to the particularly broad distribution in pore diameter in this type of sandstone~\cite{PoreGeoLinxian}, it is well suited to provide non-trivial wetting-phase geometries. We use a~subsample size of~$100^3$ voxels, which is comparable to similar studies for diffusion prediction~\cite{WANG2020100035,CNNPrifling} or permeability prediction~\cite{TANGPhyCNN,RapidEstimate} using neural networks.
\par
The sampling procedure follows the sliding frame approach as presented in~\cite{Sudakov}. As such, more samples are obtained compared to standard dissection while the subsamples can still be considered independent, cf.~\cite{RapidEstimate}. Moreover, the data set is enriched by rotations around the $y$~and $z$~axis. 
\par
Finally, impermeable (disconnected) samples are excluded from the data set as their diffusion is zero independent of the saturation level. Likewise, disconnected pore space is removed from the samples since it does not contribute to diffusive transport. In doing so, we guarantee in particular that the morphological algorithm described in \cref{sec:MorphModel} only positions the non-wetting phase in accessible parts of the pore space. Applying the above workflow, a~data set encompassing \num{5000}~samples is established.

\subsection{Morphological model}
\label{sec:MorphModel}

In the following, we present the morphological model used to compute the phase distribution of the wetting and non-wetting phase. As already illustrated in~\cite{HILPERT2001243}, well-established morphology models are capable of providing reasonable approximations of the wetting behavior in porous media. The morphological algorithm used in this paper consists of placing balls of radius $r$ within the pore space, cf.~\cite{MU201616}. The radius is derived from the Young--Laplace equation, relating capillary pressure $p_c$ and surface tension~$\gamma$ to the curvature of the separating interface in terms of its spherical radius~$r$, \ie, for zero contact angles, it holds:
\begin{align*}
    p_c=2\frac{\gamma}{r}.
\end{align*}
In order to reproduce a~virtual drainage process with gradually increasing capillary pressure, such balls are placed within the pore space with decreasing radius. This consideration leads to the well-known \emph{maximum-inscribed-sphere} methodology~\cite{Silin11} that we apply to our samples. This methodology is implemented efficiently following the approach of~\cite{MU201616}. 
Initially, the sample is fully saturated with respect to the wetting phase. In the first step, an~Euclidean distance map $d_1:[0,1]^3\to \mathbb{R}_{\geq 0}$ is computed measuring the distance of any point in the sample to the solid matrix. Matlab's built-in function \texttt{bwdist} is able to compute this function with linear complexity. The maximizer of $d_1$ corresponds to the center of the first and largest ball $\mathcal{B}_1$ to inscribe into the pore space while the related value of $d_1$ reflects its radius. Subsequent balls are now additionally required to exhibit a midpoint within the remaining wetting phase. Formally, this is achieved by setting 
\begin{align}
    d_i = \begin{cases}
    0 & \text{in } \mathcal{B}_{i-1}, \\
    d_{i-1} & \text{else}.  
    \end{cases},\quad i\geq 2,\nonumber
\end{align}
\ie deleting the last ball from $d_{i-1}$, and iterating the procedure with $d_i$. The algorithm terminates as soon as the requested saturation value $s_\mathrm{w}$ with respect to the wetting phase is undercut.

\begin{figure}[!ht]
    \centering
    \includegraphics[width=0.3\textwidth]{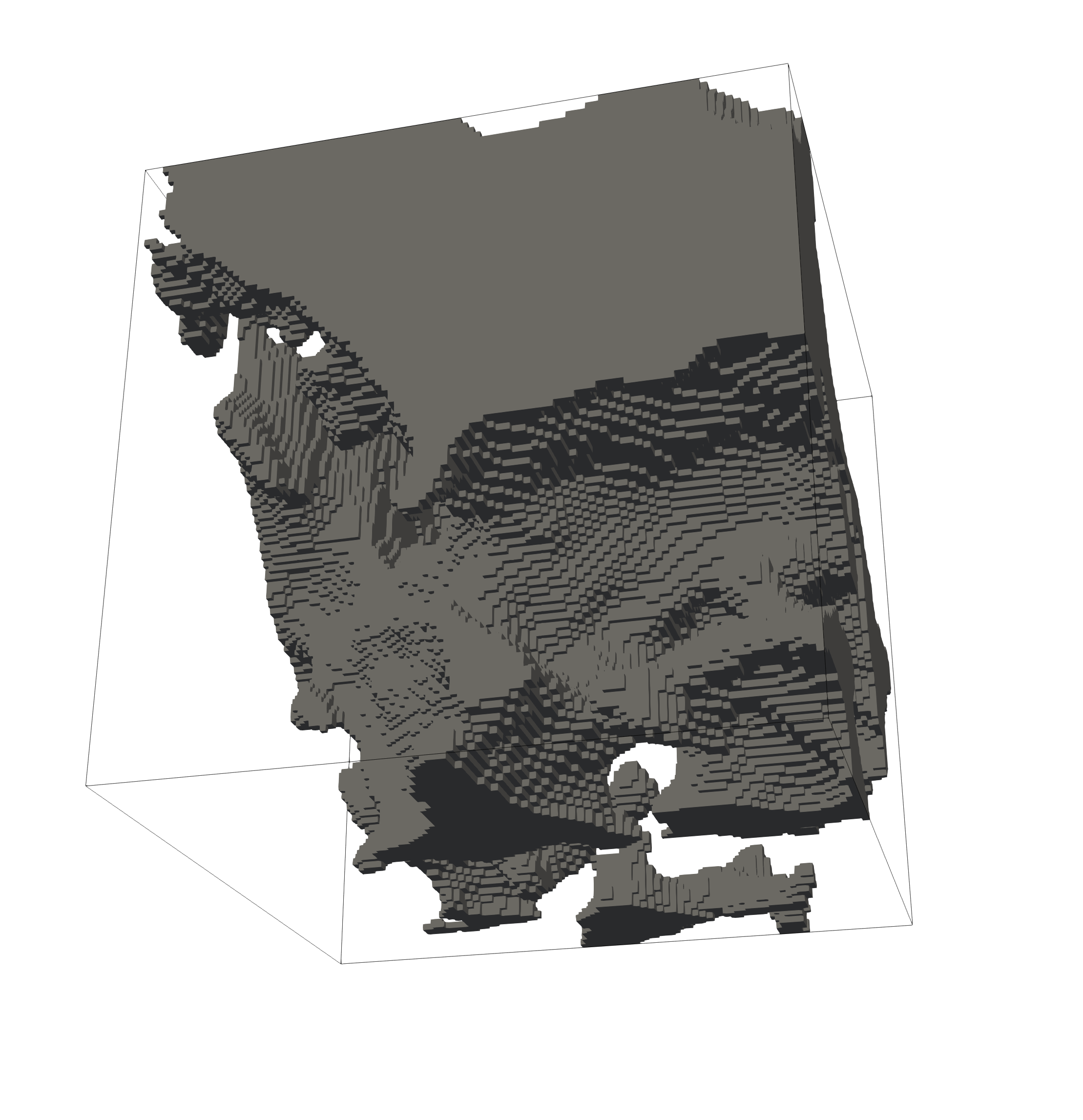}
     \includegraphics[width=0.3\textwidth]{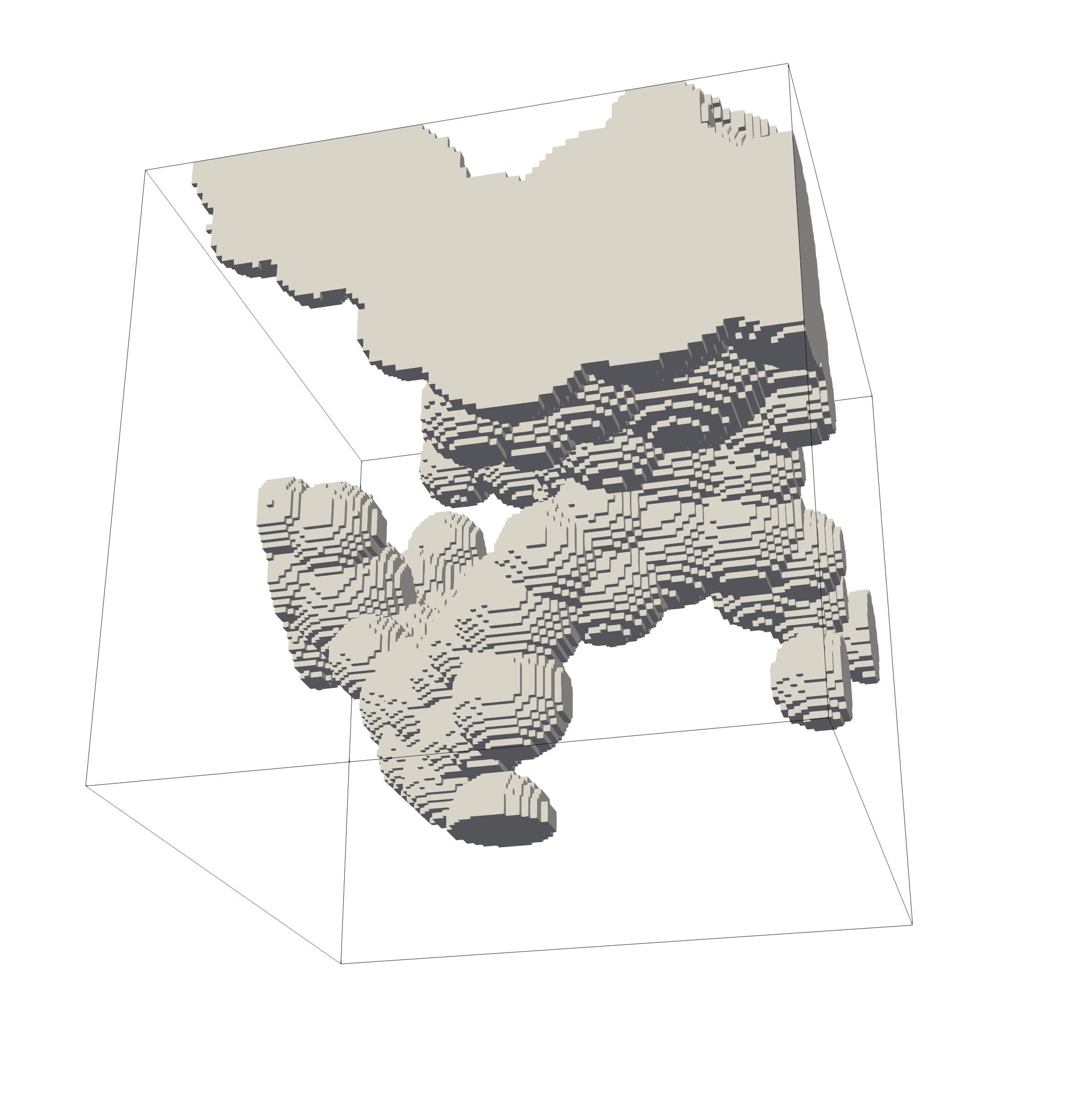}
      \includegraphics[width=0.3\textwidth]{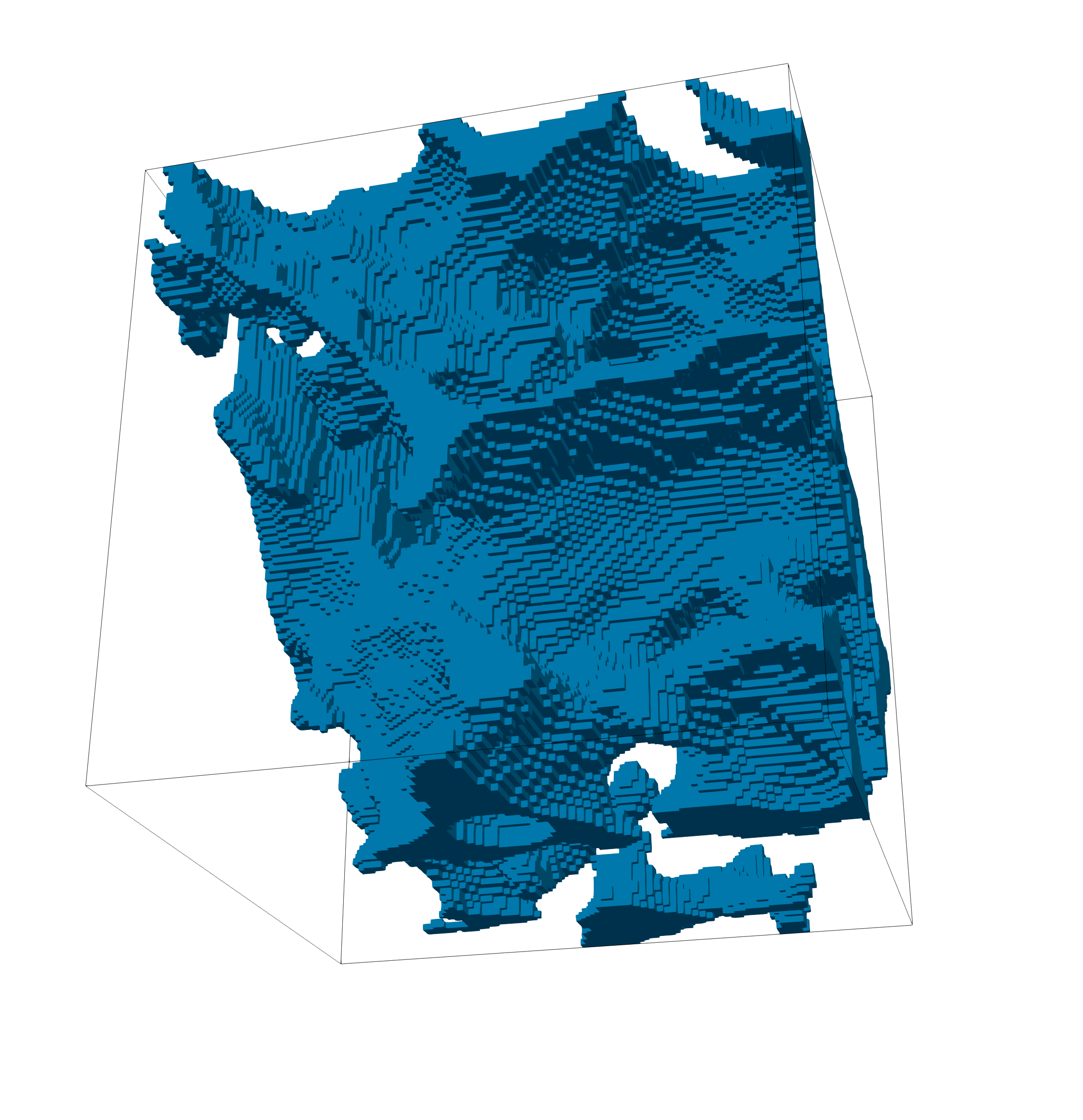}
    \caption{Illustration of morphological model for target saturation $s_\mathrm{w}=0.5$. Full pore space (left) being accordingly decomposed into a~non-wetting phase (middle) and a wetting fluid phase (right) of equal volume. In total, the non-wetting phase domain consists of $80$ overlapping spheres, ranging in radius between $20.38$ and $5.35$ voxel lengths. }
    \label{fig:Morphology}
\end{figure}

However, depending on the pore-geometry, this algorithm may lead to a~significant undershooting in $s_\mathrm{w}$, \eg in cases including macro-pores. In order to guarantee an~accurate value of $s_\mathrm{w}$, the radius $r_\mathrm{final}$ of the last sphere $\mathcal{B}_\mathrm{final}$ is tuned to adjust for the desired saturation as precisely as possible, cf.~\cite{MU201616}. The resulting subdivision of the pore space into the two fluid domains is illustrated by one of the training samples in \cref{fig:Morphology}.

\par
We subject each of the \num{5000}~samples generated as described in \cref{SEC:DataSampling} to the above morphological model for six different saturation levels with respect to the wetting phase, namely $s_\mathrm{w} \in \{1,0.9,0.8,0.7,0.6,0.5\}$. We note that using these values, the domain of the wetting phase remained connected for each considered sample. Throughout this paper, we denote the arising geometries of the wetting phase by a pair $(i,s_\mathrm{w})$ of the original sample index $i\in\{0,\dots,4999\}$ and the saturation $s_\mathrm{w}$. As such, the generated data set contains a~total of \num{30000} samples. 

\subsection{Forward simulation}
\label{SEC:ForwardSim}
To apply a~supervised learning approach as outlined in~\cref{SEC:MachineLearningModel}, each sample within the data set derived by the methods of Section~\ref{SEC:DataSampling}, \ref{sec:MorphModel} needs to be labeled with a~computed relative diffusion $D_\mathrm{cmp}$ that we use as the reference value. 
\par
To this end, in~\cref{SEC:TransportSolve}, we describe the method used for our diffusive transport simulations on the wetting-phase domain. More precisely, for each of the $\num{30\,000}$ subsamples, a~stationary concentration field and the associated flux is computed by solving~\cref{EQ:TransportProblem} using a~mixed finite element scheme.
\par
In~\cref{SEC:DiffusionEstimation}, the relative diffusion is calculated by averaging the mass flux through the sample in the direction of transport.

\subsubsection{Computation of the concentration field}
\label{SEC:TransportSolve}
We consider the stationary diffusive transport equation in mixed form
\begin{subequations}
\label{EQ:TransportProblem}
\begin{align}
-\div \vec{j}             &= 0 && \text{in}~\Omega,\label{EQ:TransportProblem:a}\\
\vec{j}&=D_0\grad c   && \text{in}~\Omega, \label{EQ:TransportProblem:b}
\end{align}
\end{subequations}
where $\Omega\subset(0,1)^3$ is the domain that consists of the union of voxels belonging to the wetting phase of one of the considered $100^3$ voxel subsamples.
In~\eqref{EQ:TransportProblem}, $c=c(x,y,z)$ denotes the concentration field, $\vec{j}=\vec{j}(x,y,z)$ the mass flux, and $D_0$ the molecular diffusion constant.
Considering diffusion with respect to the $x$~direction, we impose the following boundary conditions with the inflow boundary $\Gamma^\text{in} = \{0\}\times (0,1) \times (0,1) \cap \partial\Omega\neq \emptyset$ and outflow boundary $\Gamma^\text{out} = \{1\}\times (0,1) \times (0,1) \cap \partial\Omega\neq \emptyset$:
\begin{subequations}
    \label{eq:boundaryCond}
    \begin{align}
    c&=1, &&\Gamma^\text{in},  \\
    c&=0, && \Gamma^\text{out}, \\
    \grad c\cdot \vec{\nu}&=0, && \partial\Omega \setminus ( \Gamma^\text{in}\cup \Gamma^\text{out}),  
    \end{align}
\end{subequations} 
incorporating in particular homogeneous Neumann boundary conditions at the interior boundary which corresponds to the pore walls, cf.~\cite{TauFactor,WANG2020100035}.
Since the data preparation methodology of \cref{SEC:DataSampling} already eliminated impermeable samples, the problem is well-posed with respect to the imposed boundary conditions.

\begin{figure}[!ht]
    \centering
    \includegraphics[width = 0.75\textwidth]{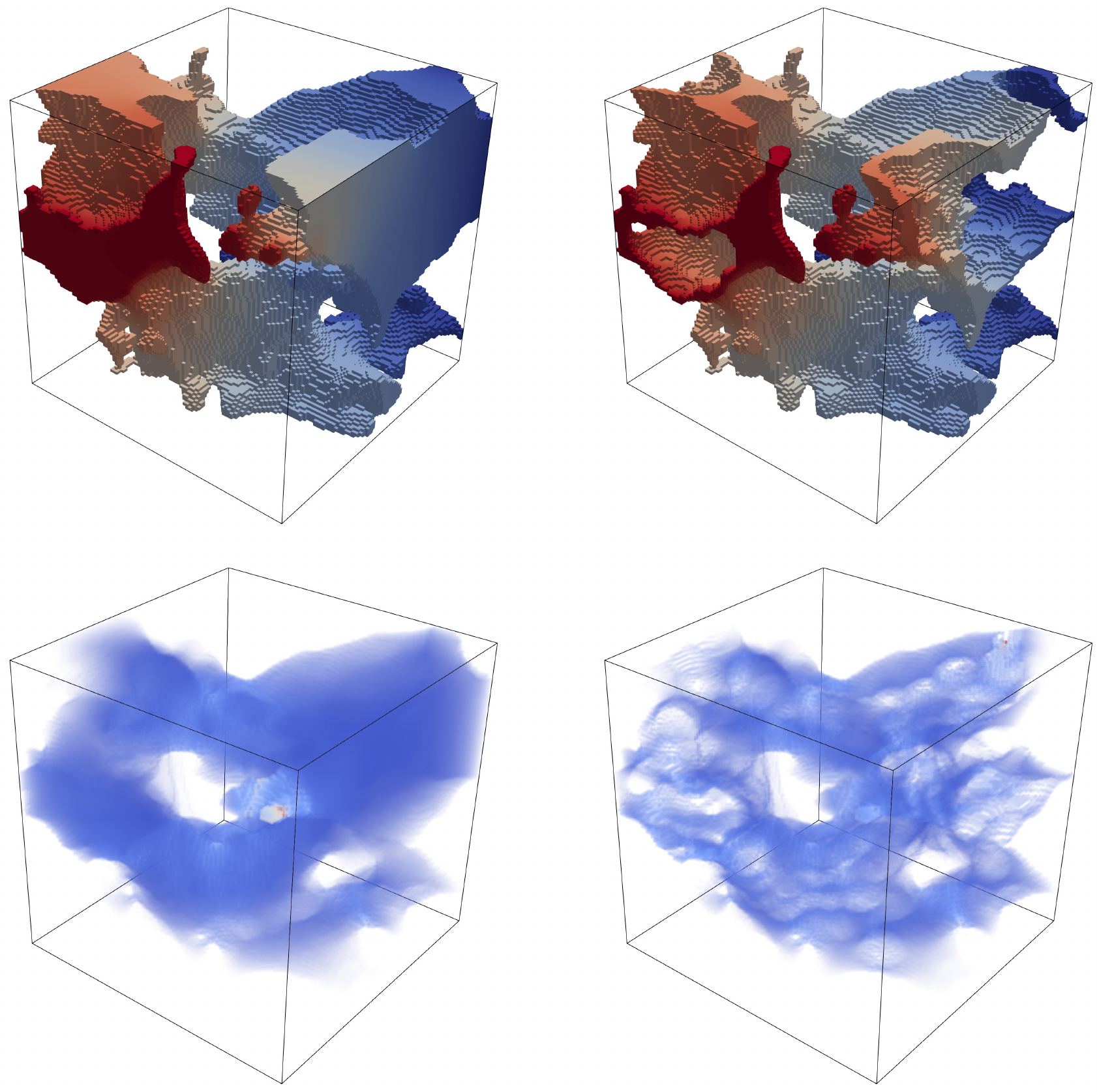}
    \caption{Simulated concentration field $c$ (top) and mass flux magnitude $\mid \vec{j} \mid$ (bottom) on a~$100^3$ voxel cube. The left column presents Sample (0,1), the right one Sample (0,0.5) corresponding to the wetting-phase distribution of Sample (0,1) at $s_\mathrm{w}=0.5$.  }
    \label{fig:Simulations}
\end{figure}

The simulations of this paper are performed using lowest-order mixed finite elements~($RT_0/P_0$). To solve the arising linear system, a~preconditioned MINRES solver is used with a~relative tolerance of $10^{-8}$. The software is implemented in the distributed-parallel framework of mfem~\cite{mfem}. As we will illustrate in Section~\ref{SEC:DiffusionEstimation}, simulations performed on an~unrefined mesh already provide sufficient accuracy for our purposes. In \cref{fig:Simulations}, we illustrate the concentration fields and diffusive fluxes for one sample at saturation levels $s_\mathrm{w}=1$ and $s_\mathrm{w}=0.5$. The effective diffusion~$D_\mathrm{cmp}$ according to~\cref{EQ:DiffTensor} requires the computation of the mass flux~$\vec{j}$, which is an explicit unknown in the mixed finite element method.

\subsubsection{Diffusion estimation and validation}
\label{SEC:DiffusionEstimation}
In order to obtain the relative diffusion of our data samples, we average the normal outward flux~$\vec{j}\cdot  \vec{\nu}$ over the outflow boundary
\begin{align}
\label{EQ:DiffTensor}
    D_\mathrm{cmp} =\frac{1}{D_0 \cdot |\Gamma^\text{out}|} \int\limits_{\Gamma^\text{out}} \vec{j}\cdot  \vec{\nu} \; \dd\sigma
\end{align}
and normalizing it to the molecular diffusion constant $D_0$, cf.~\cref{EQ:TransportProblem}, \cite{WANG2020100035}. 

For validation, we compare the results of our implementation to the software package \mbox{TauFactor}~\cite{TauFactor}. Building upon a~finite difference discretization using ghost nodes for efficiently prescribing Dirichlet boundary conditions, \mbox{TauFactor} is capable of computing diffusion from voxelized images and was also previously applied in similar studies~\cite{CNNPrifling}. 
We consider samples of different saturation levels and investigate the convergence of solutions with respect to global mesh refinement, \ie halving the discretization length with each refinement level. In Table~\ref{tab:mfemVSTaufactor}, the individual diffusion values are presented, the coarsest (rl~0) referring to the original grid resolution~$100^3$. Comparing the results of the finest resolution (rl~2), both approaches deliver relative diffusion values that differ from each other by less than $1.5\%$, cf.~\cref{tab:mfemVSTaufactor}, over the whole range of considered saturations. However, values computed by \mbox{TauFactor} seem to converge significantly slower towards the respective reference value under mesh refinement. This behavior is expected despite the agreeing theoretical order of convergence due to the explicitly discretized flow variable in our $RT_0/P_0$ finite element ansatz, resulting in a~higher approximation quality at the expense of a~larger number of unknowns. 
\begin{table}[!ht]
    \centering
    \small
    \begin{tabular}{ll|ccc|ccc}\toprule
      Sample & $s_\mathrm{w}$   & mfem rl 0 & mfem rl 1 & mfem rl 2 & TauFactor rl 0 & TauFactor rl 1 & TauFactor rl 2  \\
         \midrule
   (0,1)     & 1   &   0.133\phantom{0}        &    0.133\phantom{0}       &     0.133\phantom{0}      &       0.129\phantom{0}         &     0.132\phantom{0}          &    0.133\phantom{0}     \\ 
   (3,0.9)  & 0.9  &   0.202\phantom{0}        &    0.202\phantom{0}       &     0.202\phantom{0}      &       0.197\phantom{0}        &     0.200\phantom{0}          &    0.201\phantom{0}     \\ 
   (6,0.8)  & 0.8  &   0.0902        &    0.0906       &     0.0908      &       0.0876         &     0.0895          &    0.0905     \\ 
   (9,0.7)  &0.7       &  0.0598         &   0.0602        &   0.0604        &    0.0579            &    0.0595          &0.0601         \\  
   (12,0.6)   & 0.6    &   0.0731        &    0.0740       &     0.0745      &       0.0696         &     0.0726          &    0.0738     \\ 
   (15,0.5)  &  0.5       &    0.0267       &     0.0272      &    0.0275       &   0.0245       &      0.0264          &    0.0272       
   \\\bottomrule
    \end{tabular}
    \caption{Relative diffusion of different samples and saturation levels computed by mfem and \mbox{TauFactor} for three different refinement levels (rl). By index increments of three, we avoid considering rotated versions of the same sample, cf.~\cref{SEC:DataSampling}, \ie. different transport directions of the same geometry.  }
    \label{tab:mfemVSTaufactor}
\end{table}

Our diffusion simulations required $59.3$~MINRES~iterations on average to converge to a~relative residual of $10^{-8}$ (min: 44, max: 69). Performing two simulations in parallel each using five cores on an~Intel Xeon E5-2630~v4~CPU, all \num{30000} forward simulations required approximately $15$ hours to complete. In comparison, \mbox{TauFactor} performed the analogous calculations within two hours using the same resources. However, the results displayed in~\cref{tab:mfemVSTaufactor} indicate that mfem at refinement-level zero delivers more accurate results than the once refined TauFactor calculation, more than compensating for the increase in computation time. Moreover, the parallelism of mfem allows for strong scalability in comparison to TauFactor. Consequently, we use the mixed discretization results due to their superior computation time / accuracy ratio.

\subsection{Machine learning model}
\label{SEC:MachineLearningModel}

In the following, we introduce our CNN architecture. 
\par
In CNNs, a~given input vector $x_0\in\mathbb{R}^{n_0}, n_0 \in \mathbb{N}$ is successively propagated from one layer of processing nodes (neurons) to the subsequent one, until the final output layer (L$^\text{th}$ layer) is reached as follows: 
\begin{align}
\label{eq:SingleLayer}
    x_i = \sigma_i \left(W_i  x_{i-1} +b_i\right),
\end{align}
with $x_i\in\mathbb{R}^{n_i}$, $n_i\in \mathbb{N}$, $i\in \{0,\dots L\}$. Here, $W_i\in \IR^{n_{i},n_{i-1}}$ denote the weight matrices, $b_i\in \IR^{n_{i}}$ the so-called biases and $\sigma_i$ represent non-linear functions. While the layer structure constitutes fixed hyper-parameters which we chose as described in~\cref{TAB:CNNlayout} following~\cite{WANG2020100035}, $\bigcup_{i=1}^L W_i \cup b_i$ is the set of trainable parameters. For a~more detailed and general introduction to the field of neural networks, we refer to the literature, \eg~\cite{aggarwal2018neural,Alzubaidi2021ReviewOD}. 
\par
Specifically, convolutional layers (\texttt{conv}) are used in our CNN to exploit local correlations of the image data for feature detection. As well known from applications in image analysis, cf.~\cite{FeatureExtractionCNN}, a~single convolutional layer may only be able to extract low-level features like edges, while their successive application accesses more complex characteristics. Moreover, we chose upper layers as dense layers (\texttt{dense}) globally cross-linking the data and facilitating the final interpretation of the collected image features. Moreover, we use batch normalization layers (\texttt{BN}) to renormalize mean value and variance of the input data, increasing stability and convergence speed during learning. As an~additional building block, max-pooling layers (\texttt{MP}) are used to reduce the dimensionality. For a~more extensive background on design choices for CNNs and well-established layer types, we refer to~\cite{Alzubaidi2021ReviewOD}.   
\par
In comparison to the architecture suggested by~\cite{WANG2020100035}, we dispense with dropout layers and use instead substantial $L^2$-regularization of the weights to reduce overfitting which slightly improved performance in our setting. As such, large amplification factors are penalized by the overall cost functional improving robustness. 
\par
In~\cref{sec:CNNperformance}, CNNs as described here are trained to predict relative diffusion values from the full pore space. As the morphological model, cf.~\cref{sec:MorphModel}, explicitly entered the computation of those values, it is implicitly incorporated by the network. 

\begin{table}[!ht]
    \centering
    \begin{tabular}{l l r}
      \toprule
      \textbf{block}  & \textbf{layers} & \textbf{learnables}                               \\
      \midrule
      input1  & image input 100{$\times$}100{$\times$}100 & ---               \\
      conv1  & \texttt{conv}(32,5) -- \texttt{BN} -- \texttt{LeakyReLU}(0.1)  -- \texttt{MP}(5,5) & 4\,096 \\
      conv2  & \texttt{conv}(64,5) -- \texttt{BN} -- \texttt{LeakyReLU}(0.1) -- \texttt{MP}(4,4) & 256\,192 \\
      conv3  & \texttt{conv}(100,3) -- \texttt{BN} -- \texttt{LeakyReLU}(0.1)  & 173\,100  \\
      dense1 & \texttt{dense}(64) -- \texttt{LeakyReLU}(0.1)   & 800\,064              \\
      dense2 & \texttt{dense}(32) -- \texttt{LeakyReLU}(0.1)   & 2\,113              \\
      output & \texttt{regression}(1) &  ---  
      \\\bottomrule
      \vspace{0.05cm}
    \end{tabular}
    \caption{Layer structure of our \mbox{PhyCNN}. Nomenclature:
    \\
    \hspace*{0.1\linewidth}\begin{minipage}[t]{0.9\linewidth}
        \begin{tabbing}
          \texttt{conv}($N$,$K$): \hspace{1em} \=  convolutional layer with $N$ channels and $K\times K$ kernel size;\\
          \texttt{BN}: \> batch normalization layer;\\
          \texttt{MP}($N$,$P$):     \>maxPooling Layer, size $N$ stride $P$;\\
          \texttt{dense}($N$):   \>dense layer with $N$ neurons;\\
          \texttt{regression}($N$): \>regression layer with $N$ neurons;\\
          \texttt{LeakyReLU}($\alpha$): \>leaky rectified linear unit, slope $\alpha$ on negative inputs.
        \end{tabbing}
      \end{minipage}
    } 
    \label{TAB:CNNlayout}
  \end{table}

\section{Evaluation}
\label{SEC:Evaluation}

In this section, we perform and analyze our machine learning studies. More precisely, we evaluate the prediction quality of our CNN, cf.~\cref{SEC:MachineLearningModel}, in two different scenarios. At first, we perform the training on the fully saturated samples ($s_\mathrm{w}=1$) and corresponding labels and evaluate how the CNN generalizes to the partially saturated situation. We refer to this setup as CNN 1. In the second setup (CNN 2), we train a structurally identical CNN to perform predictions of \emph{relative} diffusion directly from the full pore space, \ie train with the labels of the partially saturated pore-space. Since our data labels are computed relying on the morphological model of~\cref{sec:MorphModel}, it is imprinted on the training data set.  As such, we evaluate the CNNs' capability to implicitly incorporate the action of the underlying morphological model.  

In order to allow for a~transparent grading of our results, we first introduce important characteristic quantities with respect to our data set.

\subsection{Evaluation metrics}
\label{SEC:EvaluationMetrics}
In the following, we display the statistic parameters $\sigma$ (standard deviation), $R^2$ (coefficient of determination) and MSE (mean-squared error).
Measuring the spread of a~data distribution, the standard deviation $\sigma$ is defined as
\begin{align*}
    \sigma = \sqrt{\frac{1}{N-1} \sum\limits_{i=1}^N \left(t_i-\bar{t} \right)^2}
\end{align*}
for $N$ real-valued quantities $t_i$. Thereby, $\bar{t}=\frac{1}{N}\sum\limits_{i=1}^N t_i$ denotes the arithmetic mean value.
\par
Moreover, the coefficient of determination 
\begin{align*}
    R^2 = 1-\frac{\sum\limits_{i=1}^N (t_i-y_i)^2}{\sum\limits_{i=1}^N (t_i-\bar{t})^2}
\end{align*}
for targets $t_i$ and corresponding predictions $y_i$ is used to quantify the degree of correlation between two data sets. Values of $R^2$ close to one indicate a~good agreement.
\par
Similarly, the mean-squared error (MSE) is an~absolute measure of accuracy given by 
\begin{align*}
\text{MSE}= \sum\limits_{i=1}^N \frac{(t_i-y_i)^2}{N}.   
\end{align*}

\subsection{Data statistics on relative diffusion}
\label{sec:DataStatistics}

\begin{figure}[ht!]
    \centering
    \includegraphics[width=\linewidth]{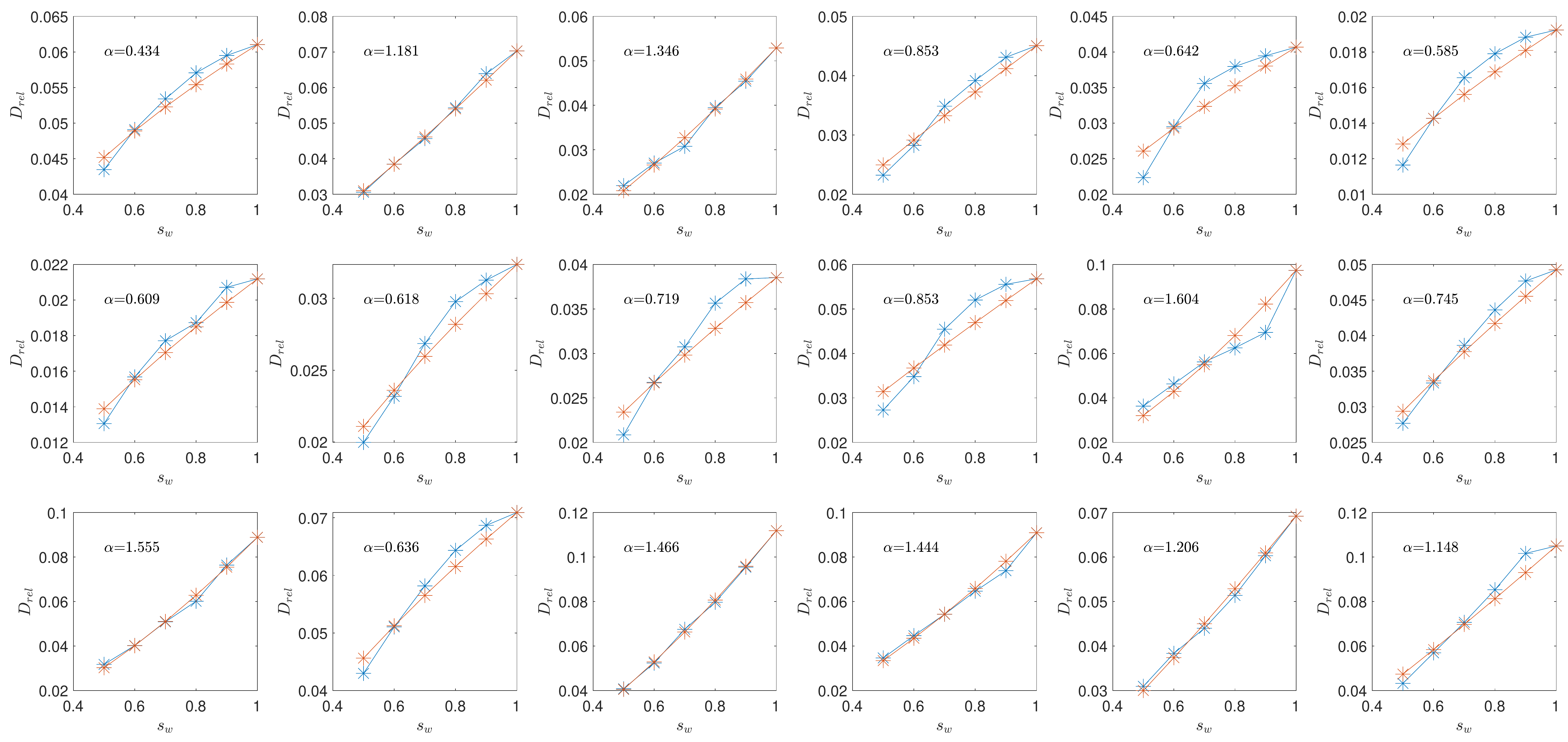}
    \caption{Relative diffusion $D_\mathrm{rel}$ over partial saturation $s_\mathrm{w}$ for 18 different samples. Blue data points refer to the computed value, red curves are related to a fit using the Bruggeman equation~\cref{eq:Bruggeman} for the displayed parameter $\alpha$. On average over all $5000$ samples, $54.7\%$ of the diffusion is maintained when restricting the available volume for diffusive transport from $s_\mathrm{w}=1$ to $s_\mathrm{w}=0.5$ with values ranging from $77.4\%$ to $2.8\%$ ($\sigma = 0.0870$). }
    \label{fig:SatCurves}
\end{figure}

In \cref{fig:SatCurves}, we present the saturation dependent relative diffusion $D_\mathrm{rel}$ of 18 randomly chosen samples. It is evident that all displayed curves show a~monotonous behavior, since an~increase of accessible pore space facilitates transport processes. However, each sample exhibits a~unique saturation--diffusion relation varying significantly over the full data set: On average over the full 5000 samples, $54.7\%$ of the original diffusion are still preserved down to a~saturation of $s_\mathrm{w}=0.5$. Thereby, the values range between $2.8\%$ and $77.4\%$ with a~standard deviation of $\sigma=0.0870$. Furthermore, we note that the $D_\mathrm{rel}/s_\mathrm{w}$-curves depicted in~\cref{fig:SatCurves} exhibit convex as well as concave behavior. These observations underpin the importance of sophisticated methods for the prediction of relative diffusion. Moreover, the data set is considered representative of a~broad class of diffusive transport scenarios in partially saturated media providing a~suitable training foundation for our CNNs outlined in~\cref{sec:CNNperformance}. 
\par
As a~well-known heuristic approach to relative diffusion, we compare our experimental findings to the Bruggeman equation \cite{TJADEN201644}: 
\begin{align}
\label{eq:Bruggeman}
    \frac{D_\mathrm{rel}}{D_\mathrm{abs}} = s_\mathrm{w}^\alpha
\end{align}
relating the relative diffusion $D_\mathrm{rel}$ to the diffusion of the fully saturated sample $D_\mathrm{abs}$. The exponent $\alpha$ is a~free parameter of the model. However, an~explicit value can be derived for special obstacle geometries such as balls ($\alpha=1.5$) and cylinders ($\alpha=2$), cf.~\cite{TJADEN201644}. Due to convexity ($\alpha>1$)/concavity ($\alpha$<1), cf.~\cref{fig:SatCurves}, a~single parameter $\alpha$ cannot approximate all curves equally well. This is reflected in the large range of optimal parameter choices $\alpha \in (0.434,1.604)$ for the individual $D_\mathrm{rel}/s_\mathrm{w}$-relations, cf.~\cref{fig:SatCurves}, again underlining the diversity displayed in the data set.     
\par
\begin{figure}[ht!]
    \centering
    \includegraphics[width=\textwidth]{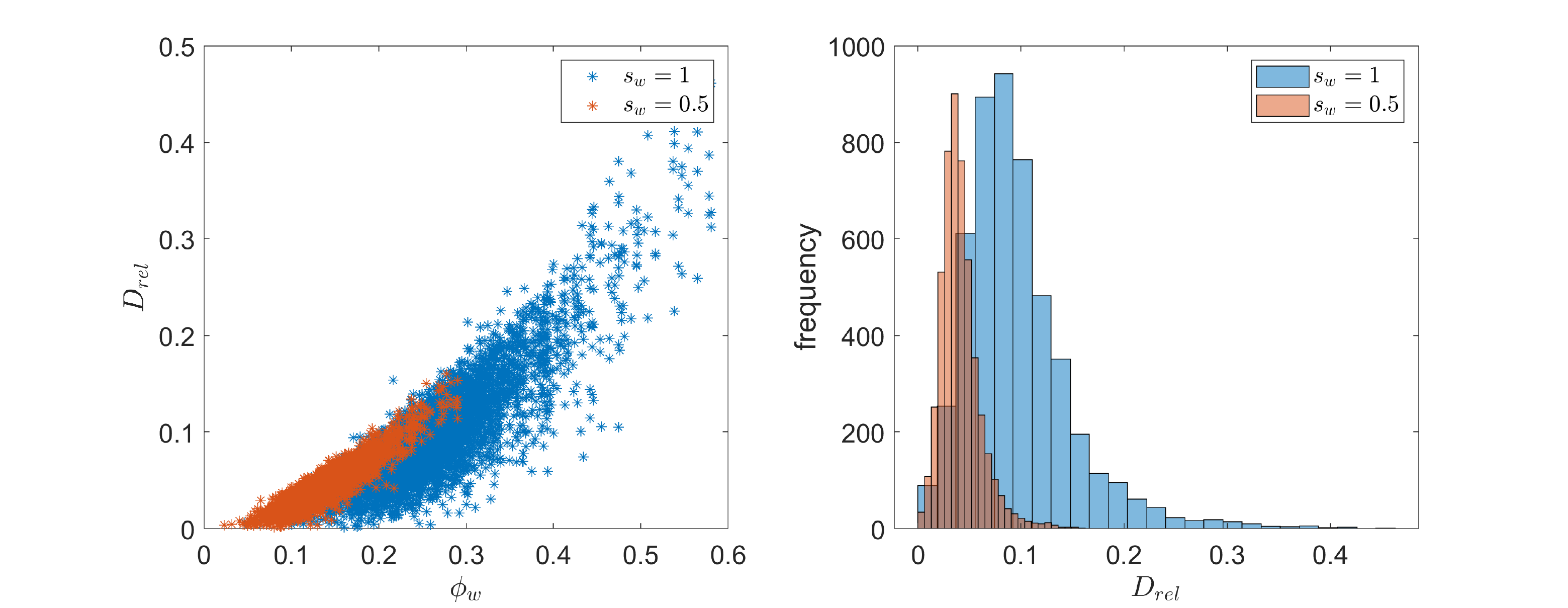}
    \caption{Correlation between relative diffusion $D_\mathrm{rel}$ and partial porosity with respect to the wetting phase $\phi_\mathrm{w}$ for samples exhibiting different saturation levels $s_\mathrm{w}$ (left). Distribution of diffusion values in a~histogram, again itemized by saturation level (right).  }
    \label{fig:Statistics}
\end{figure}

To illustrate the richness of our data set with respect to the covered porosity and diffusion range, we present further statistics in \cref{fig:Statistics}. The left image presents the dependence of the relative diffusion $D_\mathrm{rel}$ on the partial porosity $\phi_w:=s_w\cdot \phi$ of the sample with respect to the volume occupied by the wetting phase. On the right, a~histogram illustrates the distribution of $D_\mathrm{rel}$ in the generated data set. Apparently, the data is well-distributed over a~large parameter range and is therefore considered representative, again promising high generality and robustness of our CNNs. In order to regularize the data for training, data points within the sparsely sampled regions $D_\mathrm{rel}<0.01$ or $D_\mathrm{rel}>0.3$ are disregarded for CNN training and validation performed in \cref{sec:CNNperformance}, see also~\cref{fig:Statistics}. However, this procedure only discards $332$ samples in total, \ie reducing the size of our data set by approximately~$1\%$.

\subsection{CNN performance}
\label{sec:CNNperformance}

In the following, we evaluate the prediction quality of our CNNs in the setups CNN~1 and CNN~2. In both studies, the networks described in~\cref{SEC:MachineLearningModel} are trained over $13$ epochs using a~stochastic gradient descent optimizer, cf.~\cite{MATLAB:2021a}. The step size is chosen as $\eta=0.001$ initially and reduces by $60\%$ every four epochs. As a~cost functional, the mean squared error (MSE, cf.~\cref{SEC:EvaluationMetrics}) is used supplemented by an~$L^2$~regularization weighted by a~factor of $0.02$. The training procedure performed on an~Nvidia Geforce Titan~X graphics chip completed within $2.5$ hours. Throughout the study, samples were distributed among the training and validation data set using a~key of $90\% / 10\%$. The implementation is performed in Matlab R2021a~\cite{MATLAB:2021a} using the \emph{Deep Learning Toolbox}.
\par
\begin{figure}[ht!]
    \centering
    \includegraphics[width=\textwidth]{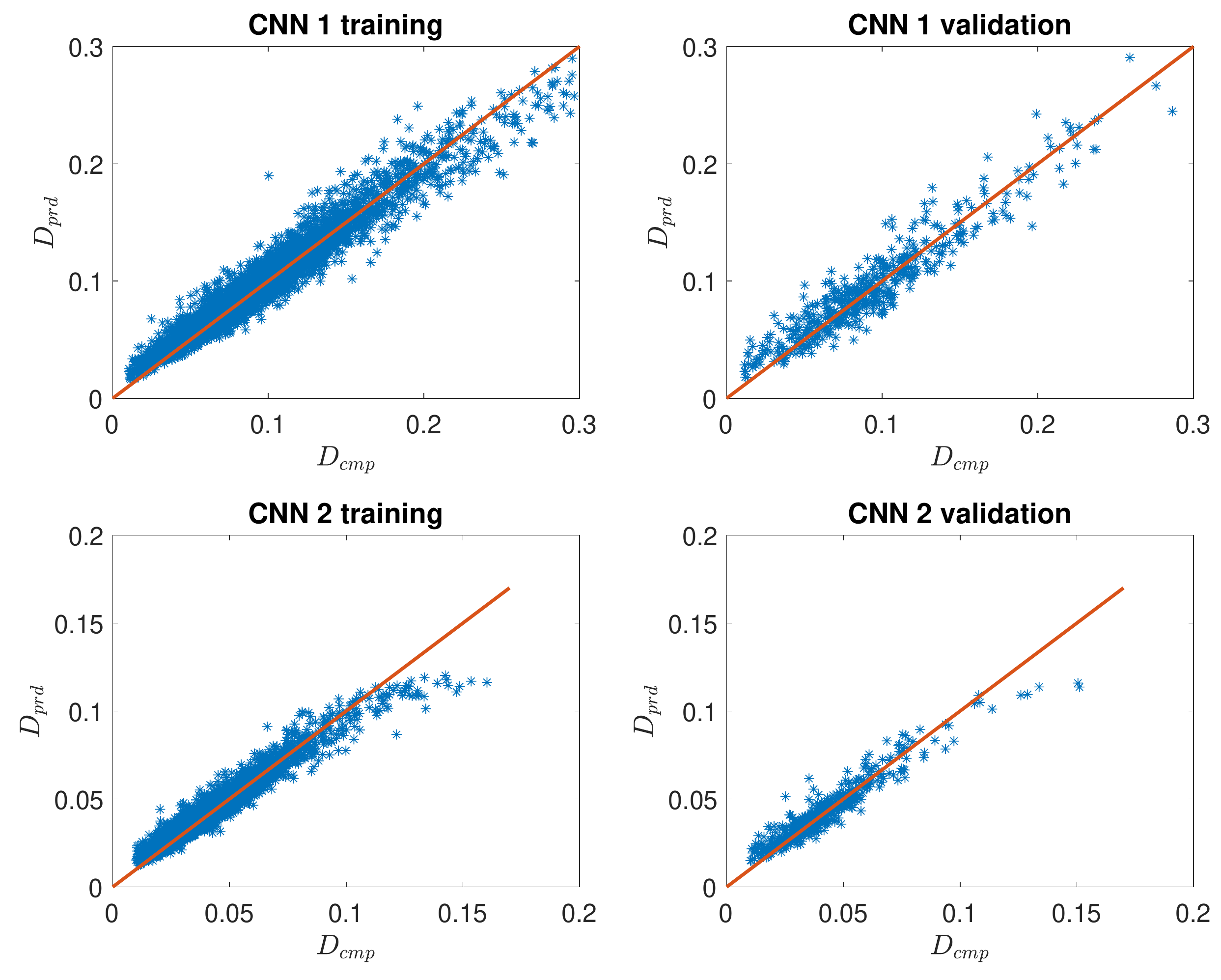}
    \caption{Accuracy of CNN predictions: Correlation plots between computed $D_\mathrm{cmp}$ and predicted relative diffusion $D_\mathrm{prd}$ for CNN 1 and CNN 2 for validation as well as training data. The red line always reflects perfect correlation.  }
    \label{fig:CNNResults}
\end{figure}

In CNN 1, the network was trained for predicting the \emph{absolute diffusion}. As illustrated in \cref{fig:CNNResults}, the network was able to perform accurate predictions on the validation data set achieving $R^2 = 89.96\%$. Comparing to $ R^2 = 93.87\%$ on the training data set, the CNN is moderately prone to overfitting. We note that the measured accuracy is lower than in the experiments~\cite{CNNPrifling,WANG2020100035} which is probably due to the use of CT-data instead of artificially generated geometries.

\begin{table}[!ht]
    \centering
    \begin{tabular}{l|ccccc}\toprule
    saturation $s_\mathrm{w}$    & $0.9$ & $0.8$ & $0.7$ & $0.6$ & $0.5$   \\
    \midrule
    $R^2$     &   $91.17\%$ & $91.31\%$ & $86.67\%$ & $69.23\%$ & $34.88\%$ 
    \\\bottomrule
    \end{tabular}
    \caption{Prediction accuracy of CNN 1 on a~validation data set for five different saturation levels $s_\mathrm{w}$. For each level, the set encompasses 500 samples independent of the training data.}
    \label{tab:Deteriorate}
\end{table}

To investigate the generalization capabilities of CNN~1 to relative diffusion, we subject the network to data samples related to partial saturations $ s_\mathrm{w} \in \{0.5,0.6,0.7,0.8,0.9\}$. More precisely, predictions were performed on 500 samples per saturation level that are derived from the validation data set for $s_\mathrm{w} =1$ and are therefore independent of the training data. The CNN is given the geometry of the wetting phase distribution as the input. As illustrated in \cref{tab:Deteriorate}, the CNN proves capable of preserving its prediction accuracy across a decent range of saturations $s_\mathrm{w}\in \{0.7,0.8,0.9,1\}$. Below these values, the geometry of the wetting phase domain seems to differ significantly from the full pore spaces in the training data, resulting in decreased precision. At $s_\mathrm{w}=0.5$, the coefficient of determination declined to $R^2 = 34.88\%$, rendering the CNN hardly usable in practice. Considering, for example, quasi-static drainage processes which naturally feature a~wide range of intermediate saturation levels, this approach requires the repeated explicit computation of the current phase distribution and subsequent prediction therefrom by a~specialized CNN. As we demonstrate in the following, both steps can be fused into a~single CNN rendering the explicit application of the morphological model superfluous.

On CNN~2, we train a~CNN of the same structure as outlined in~\cref{SEC:MachineLearningModel} to predict the relative diffusion related to a~sample with saturation $s_\mathrm{w}=0.5$ directly from the full pore-space geometry. By this second study, we assess the ability of the CNN to take the underlying morphological model (cf.~\cref{sec:MorphModel}) into account. We note that this experiment can be conducted analogously for any other saturation level. However, for $s_\mathrm{w}=0.5$, the CNN of our first study trained on full pore-space geometries only achieved unsatisfactory results, cf.~\cref{tab:Deteriorate}. As illustrated in the correlation plots of \cref{fig:CNNResults}, CNN 2 achieves a~similar performance as CNN 1 reflected in $R^2=93.22\%$ on the training data and $R^2= 89.99\%$ on the validation data despite a few outliers for particularly large relative diffusion values. Moreover, the accuracy changes only marginally when performing the predictions from the raw-samples which did not undergo the removal of disconnected pore space, cf.~\cref{SEC:DataSampling}. Therefore, samples are not required to be preprocessed in application. We note that the obtained approximation quality is significantly higher in comparison to well-established heuristic laws. First, we compare our CNN-generated results to the Bruggeman equation~\cref{eq:Bruggeman} computing relative diffusion from absolute diffusion and saturation. Given the precomputed values for $D_\mathrm{abs}$ and the target saturation $s_\mathrm{w}=0.5$, the optimal value $\alpha=1.285$ was obtained via curve fitting. Even in this specific setup (fitting for a single saturation level), the approximation quality obtained by the Bruggeman relation results in $R^2= 84.21\%$. Similar accuracy is obtained by the Millington--Quirk type relation~\cite{Millington1961PermeabilityOP}
\begin{align}
    D_\mathrm{rel} = s_\mathrm{w} ^\alpha \phi ^\beta 
\end{align}
relating saturation $s_\mathrm{w}$ and porosity $\phi$ to the target quantity via two adjustable parameters $(\alpha,\beta)$. Using the optimal values $(1.6094, 1.5340)$ obtained by curve fitting of the samples with $s_\mathrm{w}=0.5$, the approximation quality is similar to the one obtained by the heuristic law according to Bruggeman and yields $R^2=83.39\%$. However, the Millington--Quirk relation is much simpler to apply as the computation of $\phi$ is significantly cheaper than the determination of $D_\mathrm{abs}$.
In summary, we conclude that CNNs are able to perform relative diffusion predictions equally accurate as standard absolute diffusion predictions, both outpacing simple heuristic approaches. Moreover, we note that using an~Nvidia Geforce Titan X graphics chip for inferencing, CNN 2 is capable of predicting relative diffusion for a~single sample in roughly \SI{30}{ms}. Therefore, it renders large numbers of such evaluations computationally feasible as for example required in the simulation of large-scale transport phenomena in porous media with evolving phase distribution.

\section{Conclusions}
\label{SEC:SummaryandConclusions}

In this work, we investigated the feasibility of convolutional neural networks for relative diffusion prediction on partially saturated CT samples. As our study indicated, classical approaches trained on fully saturated samples corresponding to $s_\mathrm{w}=1$ only generalize well to large saturation levels (approximately $s_\mathrm{w}\geq 0.7$) and drastically lose their approximation quality for smaller values. We therefore conclude that diffusion prediction on partially saturated samples is not well-covered by existing data driven methods which moreover additionally require the explicit application of a~suitable morphological model as a~preprocessing step.    
\par
What is more, the ability of convolutional neural networks to combine the application of a~morphological model to obtain the phase distribution at a~certain saturation level with the diffusion prediction on the resulting geometry was evaluated. To do so, a~network was trained to predict the relative diffusion related to a~specific saturation level directly from the complete pore space. Our results demonstrate that CNNs are capable of performing this task as equally accurate and robust as standard diffusion prediction on fully saturated pore spaces. As such, workflows for the simulation of transport processes within partially saturated porous media can be simplified by encompassing multiple complex computations such as pore-morphology algorithms and diffusion calculation into a~single monolithic data-driven approach. The resulting speed-up specifically benefits setups that involve changes in phase distribution and geometry evolution, accordingly requiring numerous evaluations of relative diffusion over time. Moreover, in a~straight-forward approach solely relying on the results obtained in this paper, interpolation between multiple CNNs each specialized to a~certain saturation level is practicable. As such, our concepts are expected to generalize to a~continuum of saturation levels rendering them also feasible for diffusion predictions in media with dynamically (quasi-statically) changing saturation levels as occurring in drainage processes.
\par
The approach presented in this paper allows for further improvements with respect to two major aspects. First, using a~more intricate network structure, an~additional input parameter might be used to adjust the network's prediction for different contact angles between both fluid phases. As such, it generalizes the ideally non-wetting behavior of the secondary fluid investigated in this study. Second, physics-informed neural networks have demonstrated superior accuracy in the prediction of complex physical phenomena in comparison to plain data-driven approaches. For the related task of permeability assessment (in fully saturated samples), the additional supply of different pore-space characteristics such as porosity, surface area or constrictivity indices to the network have shown to improve its performance, cf.~\cite{CNNTEMBELY, CNNPerm, GaerttnerCNNPerm}. Therefore, suitably adapted pore-space quantities are expected to increase prediction accuracy in our setup as well.    

\section*{Statements and Declarations}
\subsection*{Acknowledgements}
S.~G\"arttner and A.~Meier were supported by the DFG Research Training Group 2339 Interfaces, Complex Structures, and Singular Limits.\\
N.~Ray was supported by the DFG Research Training Group~2339 Interfaces, Complex Structures, and Singular Limits and the DFG Research Unit~2179 MadSoil.\\
F.~Frank was supported by the Competence Network for Scientific High Performance Computing in Bavaria (KONWIHR).

\subsection*{Conflict of interests}
The authors have no relevant financial or non-financial interests to disclose.

\subsection*{Data availability}
The codes used in this paper will be made available within the frameworks of RTSPHEM~\cite{RTSPHEM} upon publication.
\clearpage
\bgroup
\emergencystretch 3em 
\printbibliography[heading=bibintoc]
\egroup

\end{document}